# Enhanced orbital magnetic moments in magnetic heterostructures with interface perpendicular magnetic anisotropy

Tetsuro Ueno[1,2*], Jaivardhan Sinha[1], Nobuhito Inami[2], Yasuo Takeichi[2], Seiji Mitani[1], Kanta Ono[2], and Masamitsu Hayashi[1**]

[1]National Institute for Materials Science, Tsukuba, 305-0047, Japan
[2]High Energy Accelerator Research Organization, Institute of Materials Structure Science, Tsukuba, 305-0801, Japan
*UENO.Tetsuro@nims.go.jp
**hayashi.masamitsu@nims.go.jp

## ABSTRACT

We have studied the magnetic layer thickness dependence of the orbital magnetic moment in magnetic heterostructures to identify contributions from interfaces. Three different heterostructures, Ta/CoFeB/MgO, Pt/Co/AlO$_x$ and Pt/Co/Pt, which possess significant interface contribution to the perpendicular magnetic anisotropy, are studied as model systems. X-ray magnetic circular dichroism spectroscopy is used to evaluate the relative orbital moment, i.e. the ratio of the orbital to spin moments, of the magnetic elements constituting the heterostructures. We find that the relative orbital moment of Co in Pt/Co/Pt remains constant against its thickness whereas the moment increases with decreasing Co layer thickness for Pt/Co/AlO$_x$, suggesting that a non-zero interface orbital moment exists for the latter system. For Ta/CoFeB/MgO, a non-zero interface orbital moment is found only for Fe. X-ray absorption spectra shows that a particular oxidized Co state in Pt/Co/AlO$_x$, absent in other heterosturctures, may give rise to the interface orbital moment in this system. These results show element specific contributions to the interface orbital magnetic moments in ultrathin magnetic heterostructures.

## Introduction

Perpendicularly magnetized ultrathin magnetic heterostructures are attracting great interest owing to the emergence of spin-orbit effects[1,2] which enable spin current generation, spin accumulation and formation of chiral magnetic structures[3] via strong spin-orbit coupling in the system. The strength of the spin orbit coupling (SOC) plays an important role in heterostructures with thin magnetic layer. For example, SOC at interface(s) defines the magnetic anisotropy of the system if interface contribution on the magnetic anisotropy overcomes that of the bulk.[4,5] Likewise, chiral magnetic structures emerges via the anti-symmetric exchange interaction[6,7] at interfaces[8] with large SOC. It is thus essential to understand the underlying mechanism of how SOC emerges at interfaces of ultrathin magnetic heterostructures in order to take full control of it.

Spin orbit coupling depends on the size of the spin and orbital magnetic moment as well as its coupling strength. For a typical 3$d$ transition metal, SOC is small since the orbital moment is quenched in many cases. However, this may change at interfaces where the magnetic atoms are placed in a different environment. For example, the large perpendicular magnetic anisotropy originating from a magnetic layer with an oxide interface[9,10] is considered to be a manifestation of non-zero orbital moment at the interface.[11] Identification of the presence/absense of orbital moment at interfaces and evaluation of its size, if any, is therefore important to take advantage of spin orbit effects in ultrathin magnetic heterostructures.

Here we show the magnetic layer thickness variation of the spin ($m_{\text{spin}}$) and orbital ($m_{\text{orb}}$) magnetic moments in ultrathin magnetic heterostructures to study contributions from interfaces on the orbital moment. X-ray absorption spectroscopy (XAS) and X-ray magnetic circular dichroism (XMCD) spectroscopy[12] are used to evaluate the magnetic moments in three different heterostructures: Ta/CoFeB/MgO, Pt/Co/Pt, and Pt/Co/AlO$_x$. We find that the relative orbital moment $m_{\text{orb}}$ / $m_{\text{spin}}$ increases with decreasing magnetic layer thickness when an oxide layer is placed next to the magnetic layer, suggesting an extra orbital moment forming at this interface. Interestingly, the enhancement of relative orbital moment depends on the magnetic element. For Ta/CoFeB/MgO, Fe shows an enhancement whereas Co remains constant. In contrast, Co shows a large increase when its thickness is decreased for Pt/Co/AlO$_x$. Together with the X-ray absorption spectroscopy results, we find that the degree of oxidation may play a role in defining the element specific enhancement of the relative orbital moment.

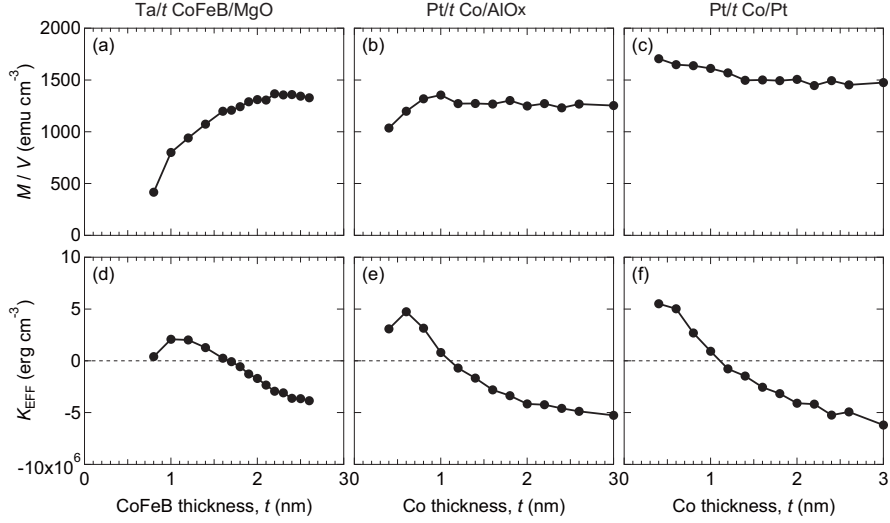

**Figure 1. Magnetic layer thickness dependence of the saturation magnetization and the magnetic anisotropy energy.**
Magnetic moment per volume $M/V$ (a-c) and the magnetic anisotropy energy $K_{eff}$ (d-f) of Ta/$t$ CoFeB/MgO (a,d), Pt/$t$ Co/AlO$_x$ (b,e) and Pt/$t$ Co/Pt (c,f) plotted as a function of magnetic layer thickness $t$. Positive $K_{eff}$ corresponds to magnetic easy axis directed along the film normal. Plots in (a) and (d) are reproduced from Ref.[17]

## Results

Magnetic properties of the films are measured at room temperature using vibrating sample magnetometry (VSM). Figures 1(a-c) show the magnetic layer thickness dependence of magnetic moment ($M$) per unit volume $M/V$. The volume ($V$) is the product of the film area ($A$) and the nominal thickness of the magnetic layer ($t$). The moment per volume $M/V$ drops significantly for decreasing magnetic layer thickness for Ta/CoFeB/MgO and slightly for Pt/Co/AlO$_x$. For Ta/CoFeB/MgO, we have previously reported that a magnetic dead layer forms and give rise to the drop in the $M/V$ for thinner CoFeB films.[13] The dead layer thickness was estimated to be ∼0.55 nm using the $x$-axis intercept value of $Mt/V$ vs. $t$. Using this method, here we find negligible magnetic dead layer for Pt/Co/AlO$_x$ and Pt/Co/Pt. Interestingly, $M/V$ increases for decreasing Co layer thickness for Pt/Co/Pt, suggesting the existence of the proximity induced magnetism in Pt.[14–16] The slope of $Mt/V$ vs. $t$ provides the saturation magnetization $M_S$ that excludes contribution from the magnetic dead layer.

The thickness dependence of the magnetic anisotropy energy $K_{EFF}$ is shown in Figs. 1(d-f). $K_{EFF}$ is obtained by calculating the integrated difference of the out of plane and in-plane field swept magnetization hysteresis loops. Positive $K_{EFF}$ corresponds to magnetic easy axis directed along the film normal. Figures 1(d-f) show that magnetic easy axis points along the film normal when the magnetic layer thickness is ∼1 nm or less. This suggests that the origin of the PMA is interface related. In order to extract the interface contribution ($K_I$) to the PMA, we fit a linear function to the $K_{EFF}t$ versus $t$ plot for the appropriate thickness range. The $y$-axis intercept of the linear fit gives $K_I$ (see Ref.[13,17] for the details of this procedure). The slope of $K_{EFF}t$ versus $t$ gives the "bulk" contribution to the magnetic anisotropy energy ($K_B$) which includes the demagnetization energy.

Table 1 summarizes the magnetic parameters for the three heterostructures studied here. In all three structures, the interface contribution ($K_I$) to the magnetic anisotropy is significant. For Pt/Co/Pt, we find $K_I$ of ∼0.6 erg/cm$^2$ per one Pt/Co interface if we assume the two interfaces contribute to the anisotropy in the same way. However, we generally find that the two interfaces possess different $K_I$ (Ref.[18] reported similar effect). Thus the contribution of the Pt/Co and Co/AlO$_x$ interfaces to the magnetic anisotropy in Pt/Co/AlO$_x$ cannot be determined uniquely.

As the three heterostructures possess significant amount of interface contribution to the magnetic anisotropy energy, these systems serve as a model system to study the presence of interfacial orbital moment. To identify its presence, we next show the analysis of the XAS and XMCD spectra measurements. Schematic configuration of the experimental setup is shown in the inset of Fig. 2(a). The X-ray incident angle is set along the film normal. An external magnetic field of 1.2 T is applied along the film normal during the measurements to measure the orbital moment perpendicular to the film plane. The spin ($m_{spin}^{eff}/n_h$) and orbital ($m_{orb}/n_h$) magnetic moments per the 3$d$ hole numbers ($n_h$) are determined using the magneto-optical sum rules[19–21] (see the Methods section for the details).

Figures 2(a) and 2(b) show typical XAS and XMCD spectra, respectively, at the $L_{2,3}$ edges of the Fe and Co atoms for

| Film structure | $M_S$ emu cm$^{-3}$ | Dead layer nm | $K_I$ erg cm$^{-2}$ | $K_B$ erg cm$^{-3}$ $\times 10^6$ |
|---|---|---|---|---|
| **1 Ta/$t$ CoFeB/2 MgO** | | | | |
| 300°C annealed | 1791 | 0.55 | 1.43 | 7.8 |
| **3 Pt/$t$ Co/1 AlO$_x$** | | | | |
| As deposited | 1233 | -0.05 | 0.88 | 1.3 |
| **3 Pt/$t$ Co/3 Pt** | | | | |
| As deposited | 1415 | -0.10 | 1.24 | 2.5 |

**Table 1. Magnetic properties of the heterostructures.** Saturation magnetization $M_S$, magnetic dead layer thickness, interface magnetic anisotropy energy $K_I$ and bulk magnetic anisotropy energy $K_B$ for the three heterostructures. Values for Ta/CoFeB/MgO are obtained from Ref.[17]

Ta/CoFeB/MgO. From the XMCD spectra, the absolute average of the $L_2$ and $L_3$ peak area provide information on the spin magnetic moment, whereas the difference is proportional to the orbital magnetic moment.

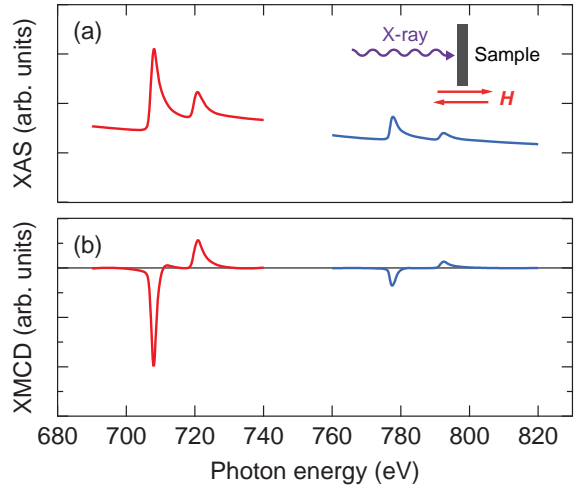

**Figure 2. Experimental setup and typical X-ray absorption and XMCD spectra.** X-ray absorption and XMCD spectra for Ta/$t$ CoFeB/MgO with $t \sim 1$ nm. Geometry of the X-ray absorption and XMCD experiments is shown in the inset of (a).

The magnetic layer thickness dependence of the effective spin ($m_{\text{spin}}^{\text{eff}}/n_h$) and orbital ($m_{\text{orb}}/n_h$) magnetic moments per hole are plotted in Figs. 3(a-c) and 3(d-f), respectively. We show the corresponding values of ($m_{\text{spin}}^{\text{eff}}/n_h$) and ($m_{\text{orb}}/n_h$) for bcc Fe and hcp Co by the dashed lines.[21] Although the CoFeB layer here is predominantly amorphous, values of bcc Fe and hcp Co can provide a rough estimate of how much the moments deviate from the corresponding bulk value. In fact, it has been recently reported that the spin and orbital moments of a single layer amorphous CoFeB (same composition, thickness $\sim$12 nm) show similar values to those of the corresponding bulk values.[22]

The spin magnetic moments per hole ($m_{\text{spin}}^{\text{eff}}/n_h$) of both Fe and Co for Ta/CoFeB/MgO (Fig. 3(a)) reduce with decreasing CoFeB thickness, similar to the trend of $M/V$ shown in Fig. 1(a). As the spin and orbital magnetic moments presented here are thickness averaged values, these results also imply the existence of the magnetic dead layer for this heterostructure. Similarly, $m_{\text{spin}}^{\text{eff}}/n_h$ of Co also decreases with decreasing Co thickness for Pt/Co/AlO$_x$ (Fig. 3(b)) and Pt/Co/Pt (Fig. 3(c)), however less rapidly compared to the drop of Ta/CoFeB/MgO. We infer the drop of $m_{\text{spin}}^{\text{eff}}/n_h$ for thin Co films is associated with the reduction in the Curie temperature as the film morphology can be more island-like for Co thickness less than $\sim$0.5 nm. The difference of the Co thickness dependence of $M/V$ and $m_{\text{spin}}^{\text{eff}}/n_h$ of Pt/Co/AlO$_x$ and Pt/Co/Pt suggests that the proximity induced magnetic moments of Pt is playing a role in defining $M/V$.

The magnetic layer thickness dependence of the orbital magnetic moments per hole ($m_{\text{orb}}/n_h$) is shown in Fig. 3(d-f). In general, $m_{\text{orb}}/n_h$ is understood to be proportional to the magnitude of $m_{\text{spin}}/n_h$.[23] However, when contributions from interfaces or surface are present, such proportionality may not hold.[24,25] For example, in Ta/CoFeB/MgO (Fig. 3(d)), we find a difference in the CoFeB layer thickness dependence of Fe and Co orbital moments. The Co orbital moment steadily decreases with decreasing CoFeB layer thickness, similar to that of the spin orbital moments. In contrast, the Fe orbital moment decays with decreasing CoFeB layer thickness in much slower way. Similarly, $m_{\text{orb}}/n_{\text{h}}$ of Co for Pt/Co/AlO$_x$ (Fig. 3(e)) shows an increase with decreasing Co layer thickness.

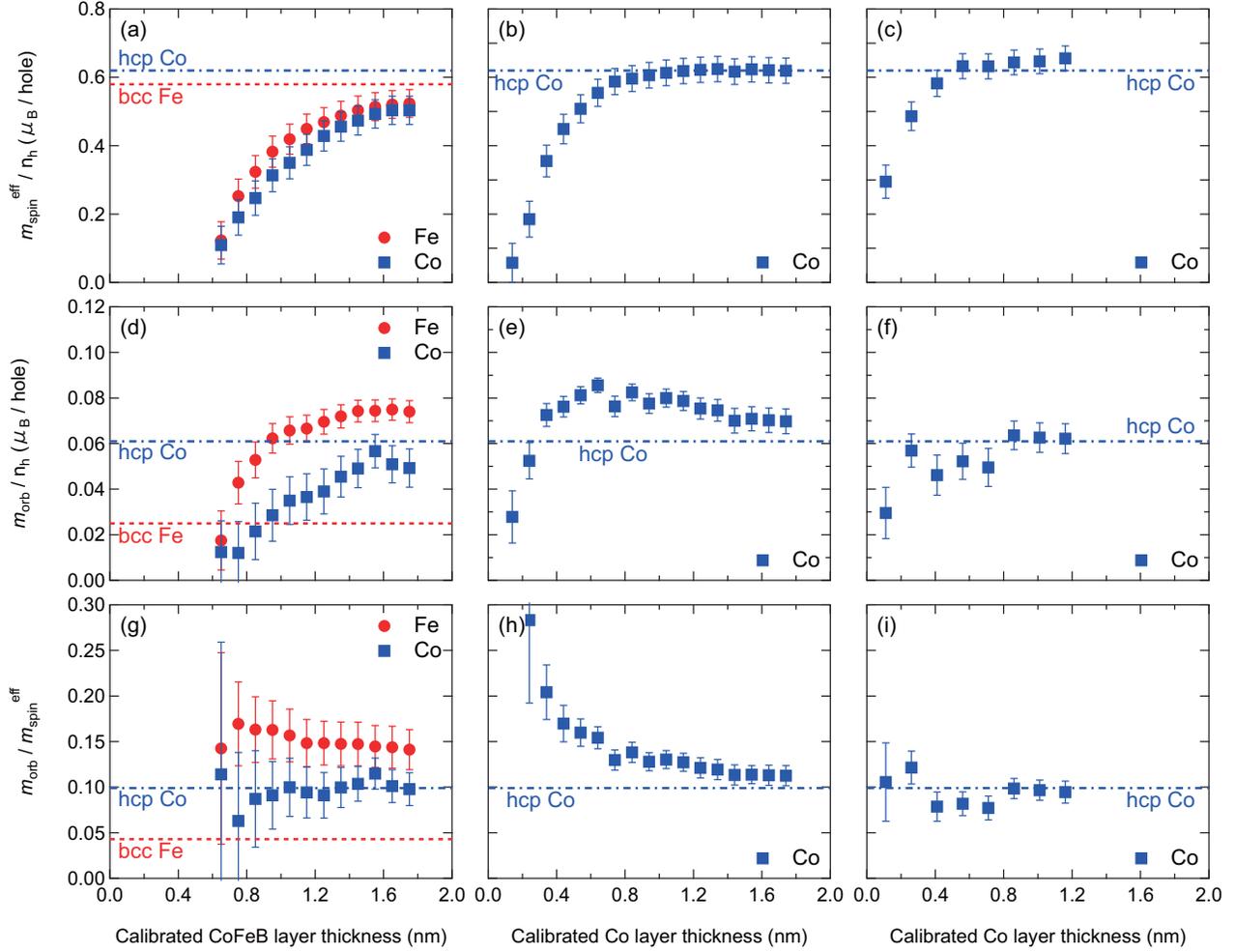

**Figure 3. Magntic layer thickness dependence of spin and orbital magnetic moments.** Spin magnetic moment per hole ($m_{\text{spin}}^{\text{eff}}/n_{\text{h}}$) (a-c), orbital magnetic moment per hole ($m_{\text{orb}}/n_{\text{h}}$) (d-f) and their ratio $m_{\text{orb}}/m_{\text{spin}}^{\text{eff}}$ (g-i) are plotted as a function of magnetic layer thickness $t$ for Fe and Co in Ta/$t$ CoFeB/MgO (a,d,g), Co in Pt/$t$ Co/AlO$_x$ (b,e,h) and Co in Pt/$t$ Co/Pt (c,f,i). See the Methods section for how the error bars are obtained. The horizontal red and blue dashed lines represent corresponding values of bulk bcc Fe and hcp Co, respectively.[21]

Since the size of the orbital magnetic moments are correlated with the spin magnetic moments in the XMCD analysis, we plot the ratio $m_{\text{orb}}/m_{\text{spin}}^{\text{eff}}$, referred to as the *relative* orbital moment hereafter, as a function of the magnetic layer thicknesses in Figs. 3(g-i). Here $m_{\text{orb}}/m_{\text{spin}}^{\text{eff}}$ for bulk bcc Fe and hcp Co are shown by the dashed lines for reference.[21] The relative orbital moment of Fe in Ta/CoFeB/MgO and Co in Pt/Co/AlO$_x$ increases with decreasing magnetic layer thickness and becomes much larger than that of the corresponding bulk values for the thinner films (similar results have been reported in Refs.[25,26]) These results suggest that the enhancement of the orbital moment originates from the interface, most likely, with the oxide layer. In contrast, the relative orbital moment of Co in Ta/CoFeB/MgO and Pt/Co/Pt is close to that of hcp Co and does not show a significant dependence on the magnetic layer thickness.

In order to identify the origin of the enhancement of the relative orbital moment for the films with the oxide layer, the X-ray absorption spectra are studied in detail. Figure 4(a-d) show the X-ray absorption spectra normalized by the $L_3$ peak

4/7

amplitude. The normalized XAS peak of both Fe and Co in Ta/CoFeB/MgO exhibits a shoulder at the high energy side (∼710 eV for Fe and ∼780 eV for Co) particularly for the thin CoFeB films. This indicates that Fe and Co atoms are partially oxidized.[27–29] In contrast, the spectra for Co in Pt/Co/Pt show little dependence on the Co thickness with almost no sign of oxidation. Interestingly, the Co XAS peak (∼777 eV) in Pt/Co/AlO$_x$ shifts to the lower energy side as the Co thickness is reduced, suggesting the existence of a satellite peak whose energy is lower than that of the pure metal state. Indeed, an oxidized Co state at this energy level has been identified previously.[27]

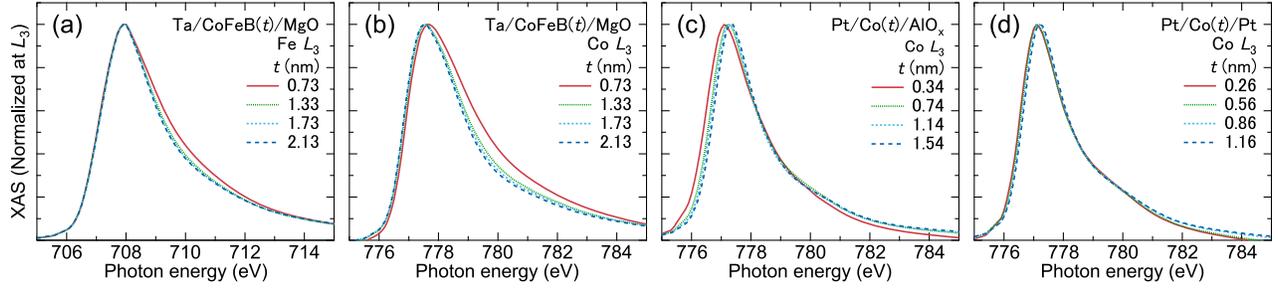

**Figure 4. Normalized X-ray absorption spectra around the $L_3$ edge for different magnetic layer thicknesses.** (a-d) X-ray absorption spectra normalized by its $L_3$ peak amplitude around the Fe (a) and Co edges (b) for Ta/$t$ CoFeB/MgO, the Co edge for Pt/$t$ Co/AlO$_x$ (c) and the Co edge for Pt/$t$ Co/Pt (d).

These results show that the electronic state of the Co atom at the oxide interface is different for Ta/CoFeB/MgO and Pt/Co/AlO$_x$. We have also studied Pt/Co/MgO and found similar results with Pt/Co/AlO$_x$. Thus the difference of the Co state between Ta/CoFeB/MgO and Pt/Co/AlO$_x$ may arise due to the texture of the magnetic layer: CoFeB is predominantly amorphous in Ta/CoFeB/MgO whereas the Co likely forms hcp in Pt/Co/oxide. Further study is required to reveal the origin of the Co state difference in the two different heterostructures.

## Conclusions

In conclusion, we have studied the magnetic layer thickness dependence of the orbital magnetic moment in magnetic heterostructures which posses large interface contribution to the perpendicular magnetic anisotropy. We find that the relative orbital magnetic moment of Fe in Ta/CoFeB/MgO and Co in Pt/Co/AlO$_x$ is enhanced for thin magnetic layer, indicating the presence of interface orbital moment for these elements. In contrast, the relative orbital moment is nearly independent of the magnetic layer thickness for Co in Ta/CoFeB/MgO and Pt/Co/Pt. From the X-ray absorption spectroscopy analysis, we find that the electronic state of Co is different for Ta/CoFeB/MgO and Pt/Co/AlO$_x$: the oxidized Co state lies at different energy level for the thin magnetic films. These results illustrate how interface orbital magnetic moment emerges in ultrathin magnetic heterostructures, which is existential for developing devices utilizing the spin orbit effects.

## Methods

**Sample preparations.** The magnetic heterostructures, Sub./1 Ta/$t$ CoFeB/2 MgO/1 Ta, Sub./3 Pt/$t$ Co/AlO$_x$ and Sub./3 Pt/$t$ Co/3 Pt (unit in nm), are deposited on 10×10 mm$^2$ Si(001) substrates using magnetron sputtering. The atomic composition of CoFeB target used for forming Ta/CoFeB/MgO is Co$_{20}$Fe$_{60}$B$_{20}$. The Ta/CoFeB/MgO heterostructures are annealed at 300°C for 1 hour in vacuum after the film deposition whereas the other structures are characterized without any post-annealing process. For the X-ray absorption and the XMCD studies, we use samples in which the thickness of the magnetic layer is varied across the substrate. We vary the beam position across the substrate to study the thickness dependence of the absorption and MCD spectra. Substrates with uniform magnetic layer thickness are used for characterizing the magnetic properties using vibrating sample magnetometry (VSM).

**XAS and XMCD experiments.** X-ray absorption spectroscopy (XAS) and XMCD experiments are performed at the undulator beamline BL-16A of the Photon Factory at the High Energy Accelerator Research Organization, Japan. X-ray absorption spectra are collected by total electron yield measuring the sample drain current. The degree of circular polarization of the incident X-rays is close to 100%. The X-ray beam size is ∼0.2 mm. Measurements are performed at room temperature in ultrahigh vacuum. The spin ($m_{spin}$) and orbital ($m_{orb}$) magnetic moments of the 3$d$ elements are determined using magneto-optical sum rules.[19–21] A linear background and an arctangent step function are subtracted from the XMCD spectra before integration for the sum rule analysis. In the spin sum rule[20] analysis, magnetic dipole moment ($m_T$) is effectively included in the spin magnetic moment ($m_{spin}^{eff} = m_{spin} + 7m_T$) since the $m_T$ term is considered to be rather small in the thickness regime of our samples.[30] The 3$d$ hole numbers ($n_h$) are needed to determine the absolute values of $m_{spin}^{eff}$ and $m_{orb}$. However, since $n_h$ is not well defined for Fe and Co in the heterostructures studied here, we show values of magnetic moments per hole ($m_{spin}^{eff}/n_h$ and $m_{orb}/n_h$). One can multiply the tabulated values[21] of $n_h$, 3.39 and 2.49 for bulk Fe and Co, respectively, to $m_{spin}^{eff}/n_h$ and $m_{orb}/n_h$ to estimate the moments per Bohr magneton.

The error bars in Fig. 3 are obtained in the following way. First the experiments for Pt/Co/AlO$_x$ is repeated two times. The difference (i.e. errors) of $m_{spin}^{eff}$ and $m_{orb}$ for the two measurements are plotted as a function of the absolute value of $m_{spin}^{eff}$ and $m_{orb}$, respectively. The error bars for $m_{spin}^{eff}$ and $m_{orb}$ in Figs. 3(a-f) are obtained by interpolating the absolute value dependence of the errors. For $m_{orb}/m_{spin}^{eff}$, the error bars are calculated using propagation of errors.

## Acknowledgements

This work was supported by the MEXT R and D Next-Generation Information Technology. The XMCD experiments were performed under the approval of the Photon Factory Program Advisory Committee (Proposal Nos. 2012V002, 2013V001, and 2014G123). The authors thank Prof. Kenta Amemiya for the support for the XAS and XMCD experiments. T.U. thank Prof. Tatsuki Oda for fruitful discussions.

## Author contributions statement

T.U. and M.H. wrote the manuscript. J.S. carried out film deposition and the VSM measurements. T.U., N.I., Y.T. and K.O. performed the XAS and XMCD experiments. All authors discussed the results and reviewed the manuscript.

## Additional information

**Competing financial interests:** The authors declare no competing financial interests.